\documentclass[aps,prb,twocolumn,superscriptaddress]{revtex4-2}

\usepackage{amssymb}
\usepackage{amsmath}
\usepackage{graphicx}
\usepackage{array}
\usepackage{multirow}
\usepackage{setspace}
\usepackage[colorlinks,linkcolor=blue,anchorcolor=blue,citecolor=red]{hyperref}

\begin{document}

\title{Electronic structures and magnetism in van der Waals flat-band material Ni$_{3}$GeTe$_{2}$}
\author{Yuanji Xu}
\email{yuanjixu@ustb.edu.cn}
\affiliation{Institute for Applied Physics and Department of Physics, University of Science and Technology Beijing, Beijing 100083, China}
\author{Xintao Jin}
\affiliation{Institute for Applied Physics and Department of Physics, University of Science and Technology Beijing, Beijing 100083, China}
\author{Haoyuan Tang}
\affiliation{Institute for Applied Physics and Department of Physics, University of Science and Technology Beijing, Beijing 100083, China}
\author{Fuyang Tian}
\affiliation{Institute for Applied Physics and Department of Physics, University of Science and Technology Beijing, Beijing 100083, China}
\date{\today}

\begin{abstract}
The study of magnetism in two-dimensional materials has garnered significant interest, driven by fundamental investigations into low-dimensional magnetic phenomena and their potential for applications in spintronic devices. Through  dynamical mean-field theory calculations, we demonstrate that Ni$_{3}$GeTe$_{2}$ exhibits flat-band characteristics resulting from the geometric frustration of its layered triangular lattice. These flat bands are further renormalized due to electronic correlation. Our calculations reveal that the magnetic order of Ni atoms is significantly influenced by both the Coulomb interaction and Hund's coupling, indicating that the physics of Ni atoms is situated in an intermediate region between Hundness and Mottness. Additionally, our results show that Ni atoms experience significant spin fluctuations in their local moments, maintaining paramagnetism at low temperatures. Furthermore, we investigate the effect of Ni vacancy, finding a substantial suppression of the density of states at the Fermi level. The physical mechanisms uncovered by our study provide a comprehensive understanding of the novel properties in this material.
\end{abstract}

\maketitle

\vspace{10pt}
\noindent        
\textbf{Introduction.}      
Two-dimensional (2D) van der Waals (vdW) materials have recently emerged as promising platforms for next-generation spintronic devices \cite{Kajale2024,Yang2022,Gong2019}. Understanding the physical mechanisms governing their diverse magnetic behaviors is crucial for both fundamental research and technological applications. A particularly intriguing issue is that materials with identical crystal structure can exhibit markedly different magnetic and macroscopic properties. For instance, Fe$_{3}$GaTe$_{2}$,  Fe$_{3}$GeTe$_{2}$ and Ni$_{3}$GeTe$_{2}$ share the same layered structure, yet display distinct magnetic characteristics. Among them, Fe$_{3}$GaTe$_{2}$ is particularly notable due to its high Curie temperature ($T_{C}$) of up to 350 K, which exceeds room temperature \cite{Zhang2022}. The other significant material is its sister compound, Fe$_{3}$GeTe$_{2}$, which has garnered considerable attention for its high ferromagnetic phase transition temperatures that persist in thin flakes, as well as its novel quantum states at low temperatures \cite{Fei2018,Deng2018,Zhu2016,Zhang2018,Zhao2021,Corasaniti2020}. For example, in Fe$_{3}$GeTe$_{2}$, ferromagnetic long-range order appears to promote heavy fermion behavior \cite{Zhu2016,Zhang2018,Zhao2021}, a phenomenon that is typically counterintuitive as heavy fermion liquids usually compete with long-range magnetic order. 

In contrast, the sister compound Ni$_{3}$GeTe$_{2}$, exhibits paramagnetism at low temperatures with a small Sommerfeld coefficient, indicating an absence of heavy fermion behavior \cite{Zhu2016,Deiseroth2006}. While extensive experimental investigations have been conducted on Fe$_{3}$GaTe$_{2}$ and  Fe$_{3}$GeTe$_{2}$, as discussed above, only limited experimental studies have been reported for Ni$_{3}$GeTe$_{2}$ to date \cite{Deiseroth2006}. Although Ni$_{3}$GeTe$_{2}$ shows no clear evidence of spontaneous magnetization, a slight increase in magnetic susceptibility is observed and obey Curie-Weiss behavior below 50 K. However, this anomaly is attributed to possible ferromagnetic impurities. The scarcity of detailed experimental data and theoretical understanding of Ni$_{3}$GeTe$_{2}$ motivates us to investigate the underlying physical mechanisms responsible for its distinct magnetic and electronic behavior.

Theoretical investigations of magnetism in this system are typically based on the Heisenberg model, with magnetic interactions estimated from density functional theory (DFT) \cite{Ruiz2024,Lee2023,Ghosh2023}. This approach has made significant progress in understanding the origins of high-$T_{C}$ \cite{Ruiz2024}. Previous calculations reveal that the 2D vdW layered materials Fe$_{3}$GaTe$_{2}$ and Fe$_{3}$GeTe$_{2}$ exhibit complex magnetic interactions, with in-plane interactions playing a crucial role in influencing the Curie temperature \cite{Ruiz2024}. However, due to the dual nature of $d$-electrons in itinerant ferromagnets, the Heisenberg model alone may not be sufficient. Indeed, studies using inelastic neutron scattering (INS) and angle-resolved photoemission spectroscopy (ARPES) have shown both itinerant and localized characteristics in Fe$_{3}$GaTe$_{2}$ and Fe$_{3}$GeTe$_{2}$, indicating that neither an itinerant nor a localized picture alone can adequately describe the magnetic properties of these systems \cite{Zhang2018,Bao2022,Wu2024,Xu2020}. Recently, Xu et al. proposed the Xu-Song-Tian model for strongly correlated itinerant ferromagnets \cite{Xu2024,Xu2025}. Their density functional theory plus dynamical mean-field theory (DFT+DMFT) calculations highlight the importance of dynamical spin fluctuations and Hund's rule coupling in accurately describing the electronic structures and magnetic properties of Fe$_{3}$GaTe$_{2}$ and Fe$_{3}$GeTe$_{2}$ \cite{Xu2024,Xu2025}. Moreover, in recent theoretical simulations on Fe$_{n}$GeTe$_{2}$ ($n=$3, 4, 5), the DFT+DMFT approach has been shown to more accurately reproduce magnetic moments, exchange interactions, and electronic properties compared to conventional GGA or GGA+U methods \cite{Ghosh2023}. Considering the research on Ni$_{3}$GeTe$_{2}$ and its distinct behavior compared to Fe$_{3}$GaTe$_{2}$ and Fe$_{3}$GeTe$_{2}$ remains limited. Therefore, elucidating the physical mechanisms underlying Ni$_{3}$GeTe$_{2}$ is essential for a comprehensive understanding of the diverse properties of these 2D vdW compounds.
  
In this work, we performed magnetic DFT+DMFT calculations to investigate the electronic structures and magnetic properties of Ni$_{3}$GeTe$_{2}$. Our results indicate that the magnetic order is strongly influenced by the Coulomb interaction $U$ and Hund's coupling $J$, suggesting that the behavior of Ni atoms lies in the intermediate region between Hund and Mott physics. Notably, our strongly correlated calculations reveal that flat bands emerging near the Fermi level originate from the geometric frustration of the triangular lattice, further renormalizing with electronic correlation. In addition, we also investigated the effect of Ni vacancy in Ni$_{3}$GeTe$_{2}$, as observed experimentally. We found a rapid decrease in the density of states (DOS) at the Fermi level, which aligns well with the experimentally measured small Sommerfeld coefficient. Our study highlights that the complex magnetic interactions, varying physical behaviors of transition-metal ions, and the chemical bonding environment collectively contribute to diverse quantum states in this type of 2D vdW compounds.

\vspace{10pt}
\noindent    
\textbf{Computational details.}   
The DFT calculations were performed using the WIEN2k package, which implements the full-potential linearized augmented plane-wave (FP-LAPW) method \cite{Blaha2023}. The Perdew–Burke–Ernzerhof (PBE) functional was employed to represent the exchange-correlation potential \cite{Perdew1996}. In the DFT+U calculations, the effective Coulomb repulsion ($U_{\rm eff}$) was varied from 3 to 7 eV, utilizing the self-interaction correction method introduced by Anisimov \cite{Anisimov1993}. Experimental crystal structures were adopted \cite{Deiseroth2006}. The muffin-tin radii were set to 2.25 a.u. for Ni, 1.99 a.u. for Ge, and 2.28 a.u. for Te. The maximum modulus for the reciprocal vector $K_{\rm max}$ was chosen such that $R_{\rm MT} \cdot K_{\rm max} = 8.0$. The k-meshes for the Brillouin zone integrations were 16$\times$16$\times$3.

To account for the dual nature of Ni-$d$ electrons, we performed DFT+DMFT calculations using the eDMFT package \cite{Haule2010,Georges1996,Kotliar2006}. In the magnetic DFT+DMFT calculations, we initialized the Ni atoms with large magnetic order by considering 10 Ni-$d$ orbitals and inducing symmetry breaking from the real part of the initial self-energy. Due to the presence of two different types of Ni atoms in Ni$_{3}$GeTe$_{2}$, two impurity solvers were employed using the hybridization expansion continuous-time quantum Monte Carlo (CT-HYB) method \cite{Werner2006,Haule2007}. The double-counting term was subtracted via the exact scheme \cite{Haule2015}. A broad range of $U$ and $J$ values were explored. We adopted $U = 5.0$ eV and $J = 0.5$ eV for systematic property calculations, which are reasonable parameters for Ni-based compounds \cite{Cao2024}. To simplify the calculations, we reduced the total electron number to account for the presence of Ni vacancy. Previous research in its sister compounds has confirmed that changing electron numbers leads to similar results in constructing supercells with vacancies \cite{Lee2023,Jang2020}. We used the maximum entropy method to perform the analytical continuation upon convergence \cite{Jarrel1996}.

 \begin{figure}
\begin{center}
\includegraphics[width=0.40\textwidth]{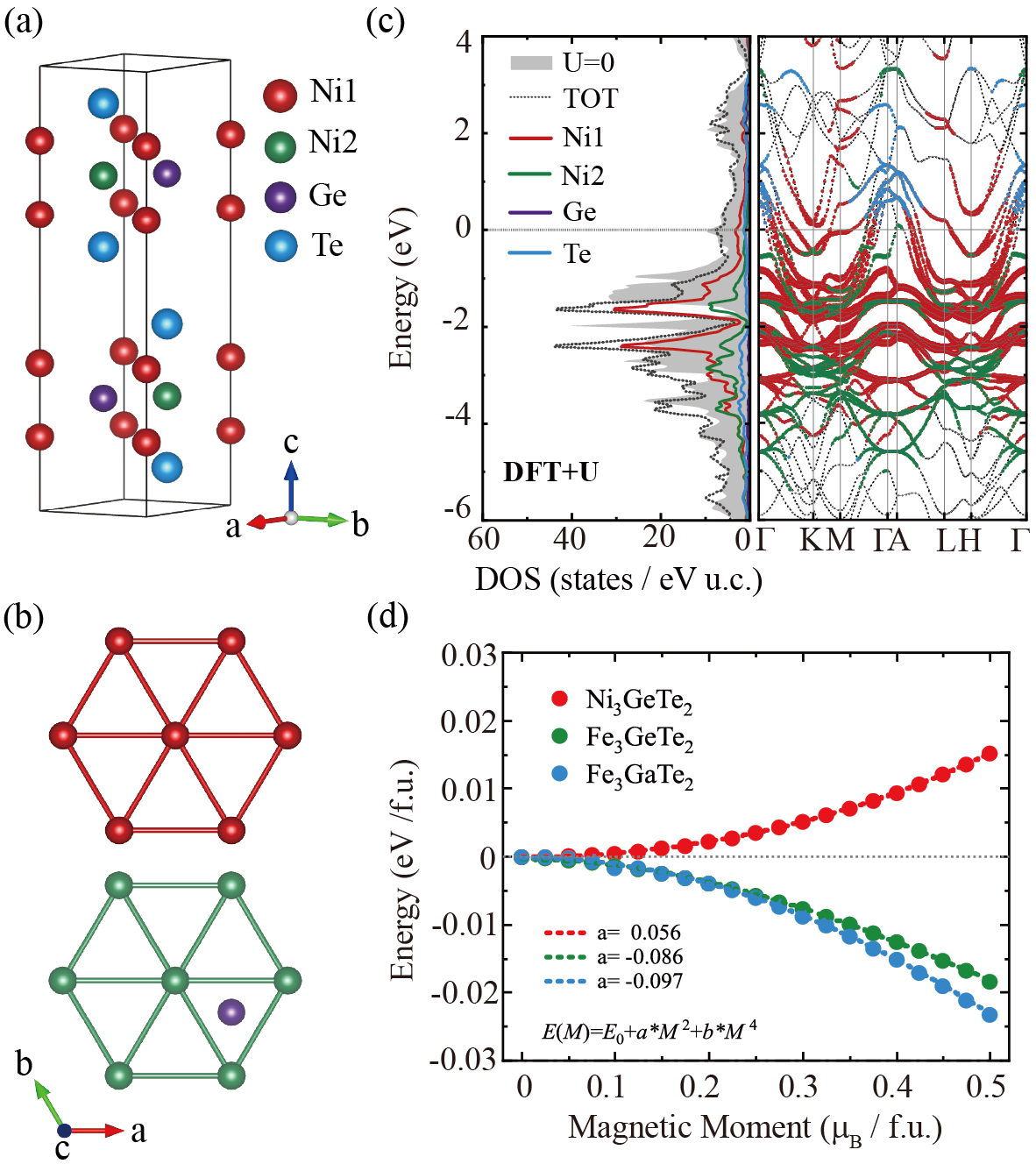}\caption{(a) Crystal structure of Ni$_{3}$GeTe$_{2}$, highlighting two distinct Wyckoff positions for Ni atoms, labeled as Ni1 (red atoms) and Ni2 (green atoms). (b) Top view of the in-plane triangular lattice formed by Ni atoms in different layers. The Ni1 layer exhibits a perfect triangular lattice, while the Ni2 layer incorporates Ge atoms in the interstitial spaces. (c) Partial density of states and electronic structure of Ni$_{3}$GeTe$_{2}$ derived from GGA+U calculations. The gray shaded region corresponds to the results obtained from pure DFT calculations. (d) Fixed spin moment DFT calculations of the total energy as a function of constrained spin magnetization for Ni$_{3}$GeTe$_{2}$. Energies are presented relative to the non-magnetic configuration, with dashed lines representing the fitting curves.}
\label{fig1}
\end{center}
\end{figure}

\vspace{10pt}
\noindent    
\textbf{Results and discussion.} 
Ni$_{3}$GeTe$_{2}$ crystallizes in a layered hexagonal structure with space group $P63/mmc$ (No. 194), featuring a substantial van der Waals gap between adjacent Te layers. Two inequivalent Wyckoff sites of Ni atoms are donated as Ni1 and Ni2, as shown in Fig.~\ref{fig1}(a). Especially, as illustrated in Fig.~\ref{fig1}(b), Ni atoms arrange into triangular lattices in the $ab$-plane, a configuration known for geometric frustration. The Ni1 layer forms a perfect triangular lattice, while the Ni2 layer has Ge atoms within this lattice's interstices. The electronic structure, obtained from DFT+U calculations with $U_{\rm eff} = 4.0$ eV, is depicted in Fig.~\ref{fig1}(c). The system converges to a non-magnetic ground state regardless of the initial magnetic configuration. Due to the higher valence electron number of Ni atoms than Fe atoms, the Ni-$d$ orbitals are more filled and pushed away from the Fermi level, resulting in a reduced DOS at the Fermi level. This behavior contrasts with that observed in the related compounds Fe$_{3}$GaTe$_{2}$ and Fe$_{3}$GeTe$_{2}$ \cite{Xu2024,Xu2025,Kim2022}. Besides, the DOS from pure DFT calculations shows a similar trend to that from DFT+U calculations, indicating weak correlation in Ni$_{3}$GeTe$_{2}$. The primary contribution to the DOS near the Fermi level comes from Ni1 atoms, while the DOS for Ni2 atoms is located at lower energy, reflecting the differing local environments of Ni1 and Ni2. The distinct behaviors associated with the two Wyckoff sites have also been reported in related compounds, highlighting the influence of local structural environments on the electronic and magnetic properties \cite{Xu2024,Xu2025,Kim2022}. Additionally, the DOS of Te atoms appears near the Fermi level, hybridizing with the Ni1 atoms. In short, the DFT+U calculations suggest that Ni$_{3}$GeTe$_{2}$ exhibits weak correlation and that Ni1 atoms predominantly contribute to its metallic properties.

\begin{table}
\caption{\label{tab1} Magnetic moments in DFT+U calculations as a function of the effective Coulomb repulsion $U_{\rm eff}$. For $U_{\rm eff} < 5$ eV, the results converge to a non-magnetic state. Conversely, when $U_{\rm eff}$ reaches 5 eV, the results converge to an ferrimagnetic state.}
\centering
\begin{spacing}{1.2}
\begin{tabular}{p{2.0cm}<{\centering}|p{2.5cm}<{\centering}|p{2.5cm}<{\centering}}
\hline
\hline
		$U_{\rm eff}$ (eV)		&		$M_{\rm Ni1}$ ($\mu_{\rm B}$)		&		 $M_{\rm Ni2}$ ($\mu_{\rm B}$)    \\ 
\hline
    0-4	&	0		& 0 		 \\
\hline
	5	&	0.4		& 	-0.09 	 \\
	6	&	0.78	& 	-0.24 	 \\
	7	&	1.0		& 	-0.5	 \\
\hline
\hline
\end{tabular}
\end{spacing}
\end{table}

To gain deeper insight into the non-magnetic order ground state observed in Ni$_{3}$GeTe$_{2}$, we calculated the Stoner parameters of this material and its sister compounds for comparison. While electron correlation effects play a crucial role in itinerant ferromagnets, the Stoner model remains widely used and is recognized as a reasonable approximation for determining the magnetic ground state \cite{Kubler2009}. Here, we performed fixed spin moment (FSM) calculations by fitting the toal energy $E$ as a function of the magnetic moment $M$ using a polynomial expansion $E(M)=E_{0}+a \times M^{2}+b \times M^{4}$, where the coefficient $a$ is related to the Stoner parameter $I$ via $a=1/N(E_{\rm F})-I$, with $N(E_{\rm F})$ representing the non-magnetic DOS at the Fermi level \cite{Kubler2009,Sieberer2006}. As shown in Fig.~\ref{fig1}(d), our fitting yields $a=$ 0.056 eV for Ni$_{3}$GeTe$_{2}$, -0.086 eV for Fe$_{3}$GeTe$_{2}$ and -0.097 eV for Fe$_{3}$GaTe$_{2}$. Using the values of $N(E_{\rm F})$ obtained from this study and previous works \cite{Xu2024,Xu2025}, the Stoner parameter $I$ is determined to be 0.16 eV (Ni$_{3}$GeTe$_{2}$), 0.38 eV (Fe$_{3}$GeTe$_{2}$) and 0.1586 eV (Fe$_{3}$GaTe$_{2}$). Consequently, the Stoner's criterion, given by $I \times N(E_{\rm F})$, is found to be 0.74 ($<$1) for Ni$_{3}$GeTe$_{2}$, 1.29 for Fe$_{3}$GeTe$_{2}$ and 2.56 for Fe$_{3}$GaTe$_{2}$. These results indicate that, based on the Stoner model, Fe$_{3}$GeTe$_{2}$ and Fe$_{3}$GaTe$_{2}$ exhibit a strong tendency toward ferromagnetism. In contrast, Ni$_{3}$GeTe$_{2}$ remains in a non-magnetic ground state. Our results indicate that Stoner's criterion can serve as a reasonable qualitative indicator in these systems.

\begin{figure}
\begin{center}
\includegraphics[width=0.43\textwidth]{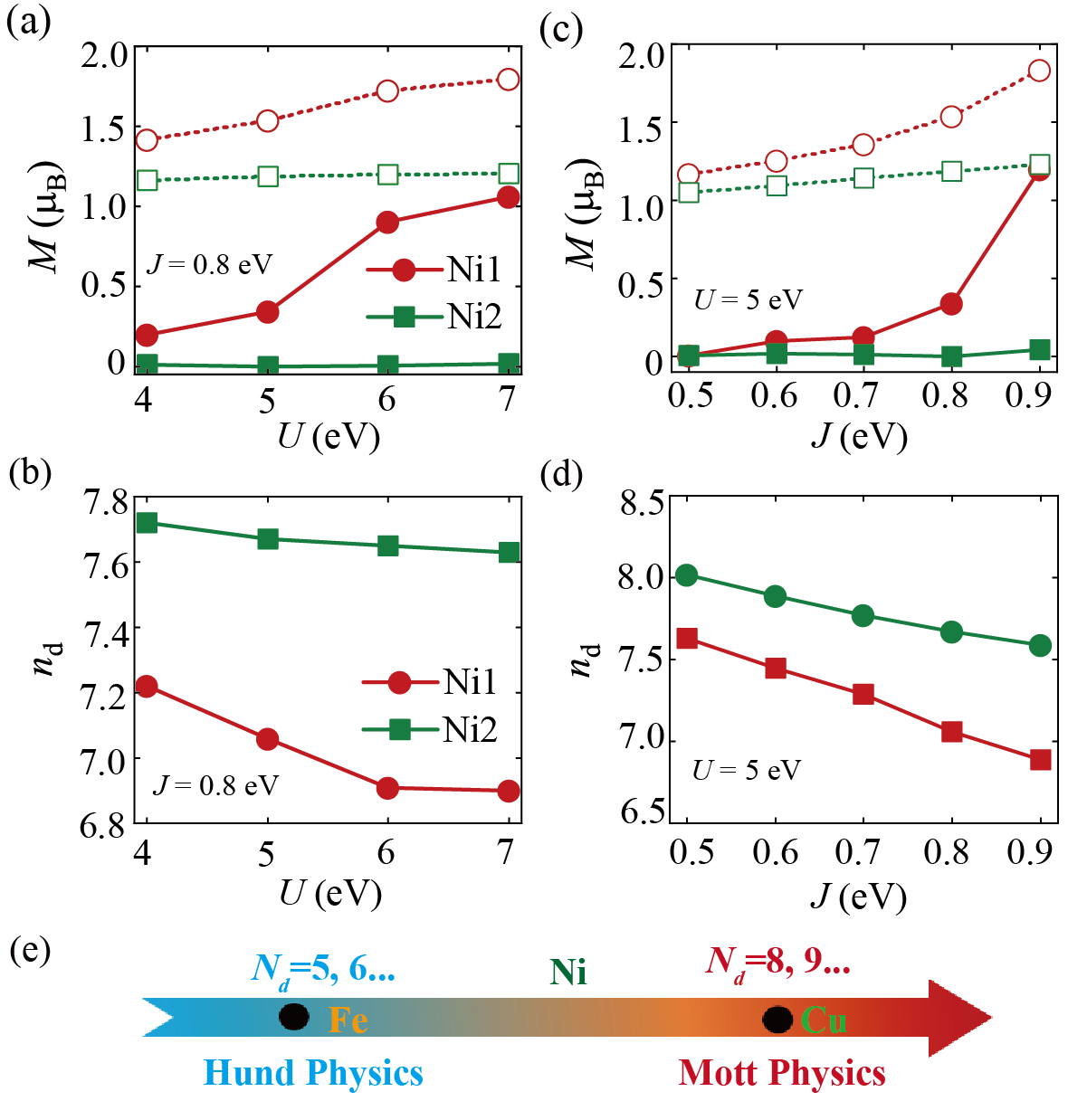}
\caption{(a) Magnetic moments (filled symbols) of Ni1 and Ni2 atoms as a function of Coulomb interaction $U$ with a fixed Hund’s coupling interaction $J$ in DFT+DMFT calculations. For comparison, the corresponding local moments of Ni1 and Ni2 atoms are represented by empty symbols. (b) Electron occupation of Ni-$d$ orbitals as a function of Coulomb interaction $U$. (c) Magnetic moments (filled symbols) of Ni1 and Ni2 atoms as a function of Hund’s coupling interaction $J$ with a fixed Coulomb interaction $U$ in DFT+DMFT calculations. The corresponding local moments of Ni1 and Ni2 atoms are shown using empty symbols for comparison. (d) Electron occupation of Ni-$d$ orbitals as a function of Hund’s coupling interaction $J$. (e) Schematic illustration of the relationship between Hund and Mott physics in relation to $d$-electron occupation.}
\label{fig2}
\end{center}
\end{figure}

In its sister compounds, Fe$_{3}$GaTe$_{2}$ and Fe$_{3}$GeTe$_{2}$, strong electronic correlation has been demonstrated to play a critical role in determining their physical properties \cite{Zhu2016,Zhang2018,Zhao2021,Xu2024,Xu2025}. To investigate the relationship between electronic correlation and magnetic ordering in Ni$_{3}$GeTe$_{2}$, we examined the magnetic states as a function of the effective Coulomb repulsion $U_{\rm eff}$ using DFT+U calculations. The results are summarized in Table~\ref{tab1}. For small values of $U_{\rm eff}$, the system consistently converges to a non-magnetic state. However, as $U_{\rm eff}$ increases to 5 eV, the system undergoes a transition to an ferrimagnetic state. The magnetic moment of Ni atoms increases significantly with increasing $U_{\rm eff}$, underscoring the intricate nature of magnetic interactions and their strong dependence on correlation effects and electronic structure \cite{Ruiz2024,Lee2023,Ghosh2023}. Similar complex magnetic interactions and multiple magnetic ground states have been observed in related compounds, such as Fe$_{3}$GaTe$_{2}$ and Fe$_{3}$GeTe$_{2}$ \cite{Ruiz2024,Lee2023,Jang2020,Yi2017}. Furthermore, our results reveal that the magnetic moments of Ni2 atoms are considerably smaller than those of Ni1 atoms, which is expected, given that Ni2 atoms have a higher electronic occupation and are more than half-filled.

To further account for dynamical spin fluctuations and the dual nature of Ni-$d$ electrons beyond the DFT method, we performed finite temperature magnetic DFT+DMFT calculations. Previous studies on its sister compounds Fe$_{3}$GeTe$_{2}$ and Fe$_{3}$GaTe$_{2}$ have demonstrated that these calculations accurately capture electronic structures, including spectral weight transfer during magnetic phase transitions and renormalized flat bands near the Fermi level \cite{Xu2024,Xu2025}, in agreement with experimental observations \cite{Wu2024,Xu2020}. As discussed above, the magnetism of Ni$_{3}$GeTe$_{2}$ exhibits distinct characteristics compared to its sister compounds, with complex behavior influenced by electronic correlation effects. Figure~\ref{fig2} presents the variation of magnetic moments for Ni1 and Ni2 atoms as functions of Coulomb repulsion $U$ and Hund's coupling interaction $J$ in DFT+DMFT calculations. Our strongly correlated calculations indicate that the magnetic moments are significantly influenced by both $U$ and $J$ interactions. Specifically, the magnetic moments of Ni1 atoms increase markedly when $U$ exceeds 4 eV, as shown in Fig.~\ref{fig2}(a). Similarly, Fig.~\ref{fig2}(c) illustrates a substantial enhancement in the magnetic moments of Ni1 atoms with increasing $J$, particularly when $J$ reaches 0.8 eV. Furthermore, the electron occupation of Ni-$d$ orbitals is closely correlated with the magnetic moments, as depicted in Fig.~\ref{fig2}(b) and (d), where a decrease in electron occupation accompanies an increase in the magnetic moments of Ni atoms. Due to the higher electron occupation of Ni2 atoms in the DFT+DMFT calculations, Ni2 atoms exhibit only weak magnetic moments. Lastly, it is important to note that the appropriate Coulomb interaction $U$ for Ni-based compounds is typically less than 5 eV \cite{Cao2024}. Given that experimental studies report no spontaneous magnetization in Ni$_{3}$GeTe$_{2}$, we adopt $U = 5$ eV and $J = 0.5$ eV for the subsequent DFT+DMFT calculations in this work.

\begin{figure}
\begin{center}
\includegraphics[width=0.43\textwidth]{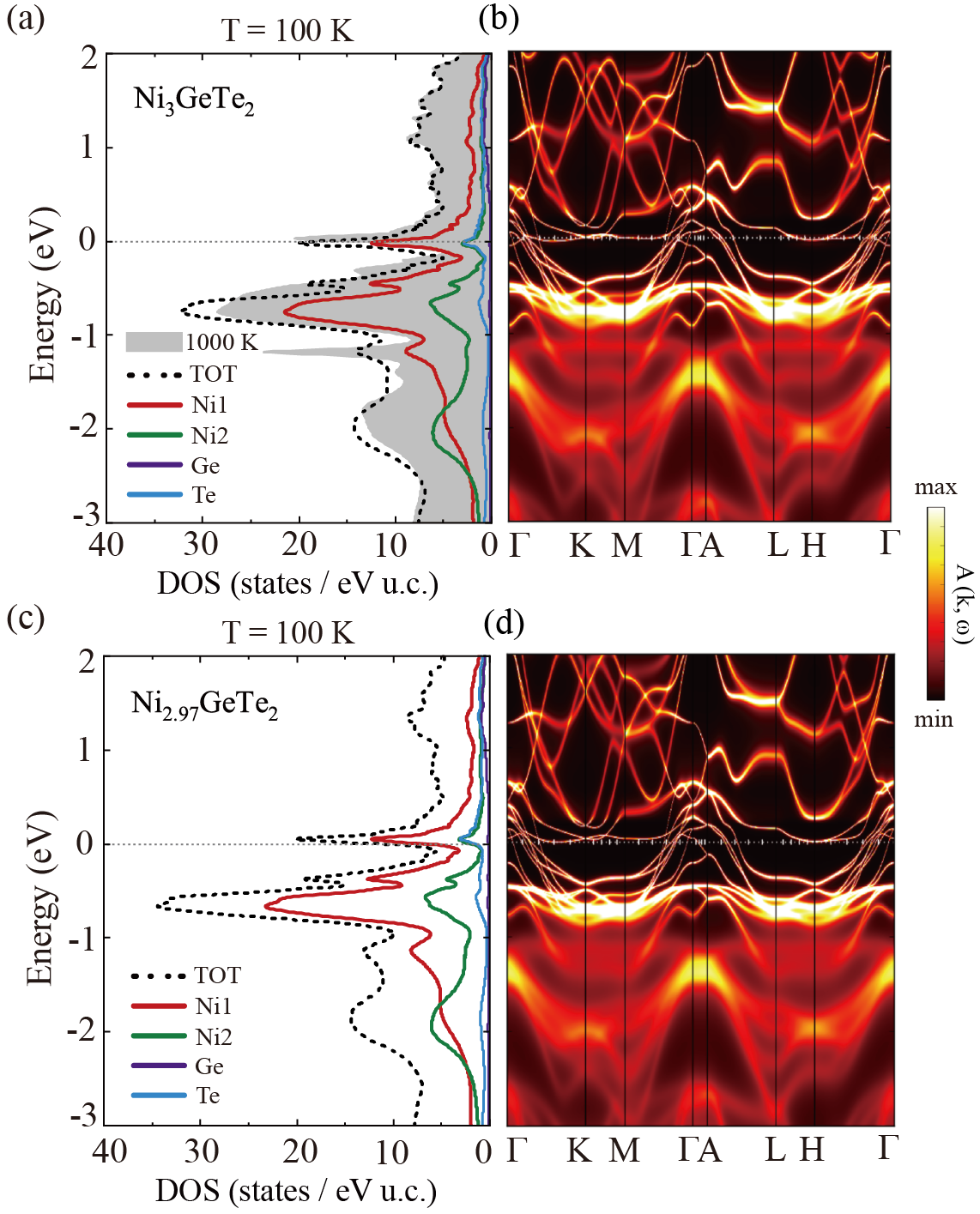}
\caption{(a) The partial density of states of Ni$_{3}$GeTe$_{2}$ at 100 K from DFT+DMFT calculations with $U = 5$ eV and $J = 0.5$ eV. The gray area represents the total density of states at 1000 K. (b) The corresponding spectral function of Ni$_{3}$GeTe$_{2}$ obtained from DFT+DMFT calculations. (c) The partial density of states of Ni$_{2.97}$GeTe$_{2}$ at 100 K from DFT+DMFT calculations, incorporating the effects of vacancies in the system. (d) The corresponding spectral function of Ni$_{2.97}$GeTe$_{2}$ at 100 K obtained from DFT+DMFT calculations.}
\label{fig3}
\end{center}
\end{figure}

We observe that, despite the absence of long-range magnetic order at low correlation strengths, the local moments of Ni1 and Ni2 atoms remain significantly large, as shown in Fig.~\ref{fig2}(a) and (c). This means that in the paramagnetic phase with zero magnetic moment, local moments can still exist at Ni atoms with random orientations. These local moments are calculated using the impurity solver within the DFT+DMFT framework. As the Coulomb interaction $U$ and Hund’s coupling interaction $J$ increase, the local moments of Ni1 and Ni2 atoms are further enhanced. Moreover, the local moments consistently exceed the magnitude of the long-range magnetic order, resulting in substantial spin fluctuations in Ni$_{3}$GeTe$_{2}$. In addition, in $d$-electron systems, the electronic occupation significantly influences the dominant correlation mechanism. When the $d$-electron occupancy is near half-filling, the system is typically sensitive to or even dominated by Hund’s coupling interaction $J$, the phenomenon often referred to as Hundness, as observed in Fe-based superconductors \cite{Yin2011}. Conversely, when the $d$-electron occupancy is close to full-filling, the system becomes more sensitive to the Coulomb interaction $U$, a regime commonly known as Mottness, as exemplified by cuprate superconductors. This conceptual framework is illustrated in Fig.~\ref{fig2}(e). Here, in the case of Ni$_{3}$GeTe$_{2}$, our findings suggest that it resides in an intermediate regime between Hund and Mott physics, as evidenced by the strong dependence of its magnetic moments on both $U$ and $J$. Such behavior of Ni-$d$ electrons has also been reported in recently discovered Ni-based superconductors \cite{Cao2024}.

Figure~\ref{fig3}(a) illustrates the partial density of states of Ni$_{3}$GeTe$_{2}$ at 100 K obtained from DFT+DMFT calculations, where the system converges to a non-magnetic state. The broad peaks below the Fermi level in the DOS are associated with the correlated Ni1 and Ni2 atoms and indicate spin fluctuations. These broad peak characters are also evident in the spectral function, as shown in Fig.~\ref{fig3}(b), which features large blurred regions extending from -3 eV to -1 eV. Additionally, the electronic structures near the Fermi level is significantly renormalized compared to DFT calculations. Notably, the renormalized bands exhibit a sharp peak at the Fermi level. We recognize these flat bands are likely a result of the geometric frustration associated with the triangular lattice formed by Ni atoms and are further renormalized due to the correlation effects of Ni-$d$ orbitals \cite{Rosa2024,Checkelsky2024}. In geometrically frustrated lattices, flat bands and multiple van Hove singularities (VHS) are frequently observed and are considered key electronic features arising from the underlying geometric frustration \cite{Hu2022,Huang2024}. Notably, even in our pure DFT calculations without any explicit treatment of electronic correlations, clear signatures of VHS and flat bands are evident (see in supplementary information), highlighting the intrinsic origin of geometric frustration in this system. And the DFT+U calculations give similar results. Moreover, the observation that Ni1 atoms, which form a perfect triangular lattice without Ge or Te atoms in the interstitial spaces, show a much higher DOS at the Fermi level than Ni2 atoms, indicating that the flat bands arise from geometric frustration. Additionally, the persistence of these flat bands at high temperatures, such as 1000 K and above, further supports the influence of geometric frustration on the formation of flat bands.

The value of the DOS at the Fermi level indicates a Sommerfeld coefficient $\gamma \approx 24$ mJ/mol K$^2$ in Ni$_{3}$GeTe$_{2}$, which is higher than the experimental result $\gamma_{\rm exp} \approx 9$ mJ/mol K$^2$ \cite{Zhu2016}. The pronounced DOS at the Fermi level is due to flat bands and is susceptible to perturbations from impurities or vacancies. Previous studies have shown that Fe$_{3-\delta}$GeTe$_{2}$, with $\delta$ ranging from 0 to 0.3, exhibits significant effects on electronic structures and magnetic properties due to Fe2 vacancies \cite{Verchenko2015}. Similarly, Ni$_{3}$GeTe$_{2}$ has reported Ni site vacancies with $\delta \approx 0.05$ \cite{Deiseroth2006}. To assess the impact of Ni vacancy, additional hole-doped DFT+DMFT calculations were performed. Figures~\ref{fig3}(c) and (d) demonstrate substantial changes in the electronic structure for Ni$_{2.97}$GeTe$_{2}$. A small amount of hole-doping results in a noticeable shift of the Fermi level towards the dip in Ni peaks, leading to a rapid decrease in the DOS at the Fermi level. Given the dependence of the magnetic moment on correlation parameters and the sensitivity of the DOS near the Fermi level to vacancies, it is plausible that Ni$_{3}$GeTe$_{2}$ could be manipulated by external parameters such as stress or pressure to achieve ferromagnetic ordering in high-quality samples. Moreover, we acknowledge the sensitivity of the DOS to the position of the Fermi level. Various types of defects or doping strategies that alter the electron count in the system could be employed to modulate the physical properties of Ni$_{3}$GeTe$_{2}$. A more comprehensive investigation of these possibilities is beyond the scope of the present work but would be a valuable direction for future studies.

\begin{figure}
\begin{center}
\includegraphics[width=0.43\textwidth]{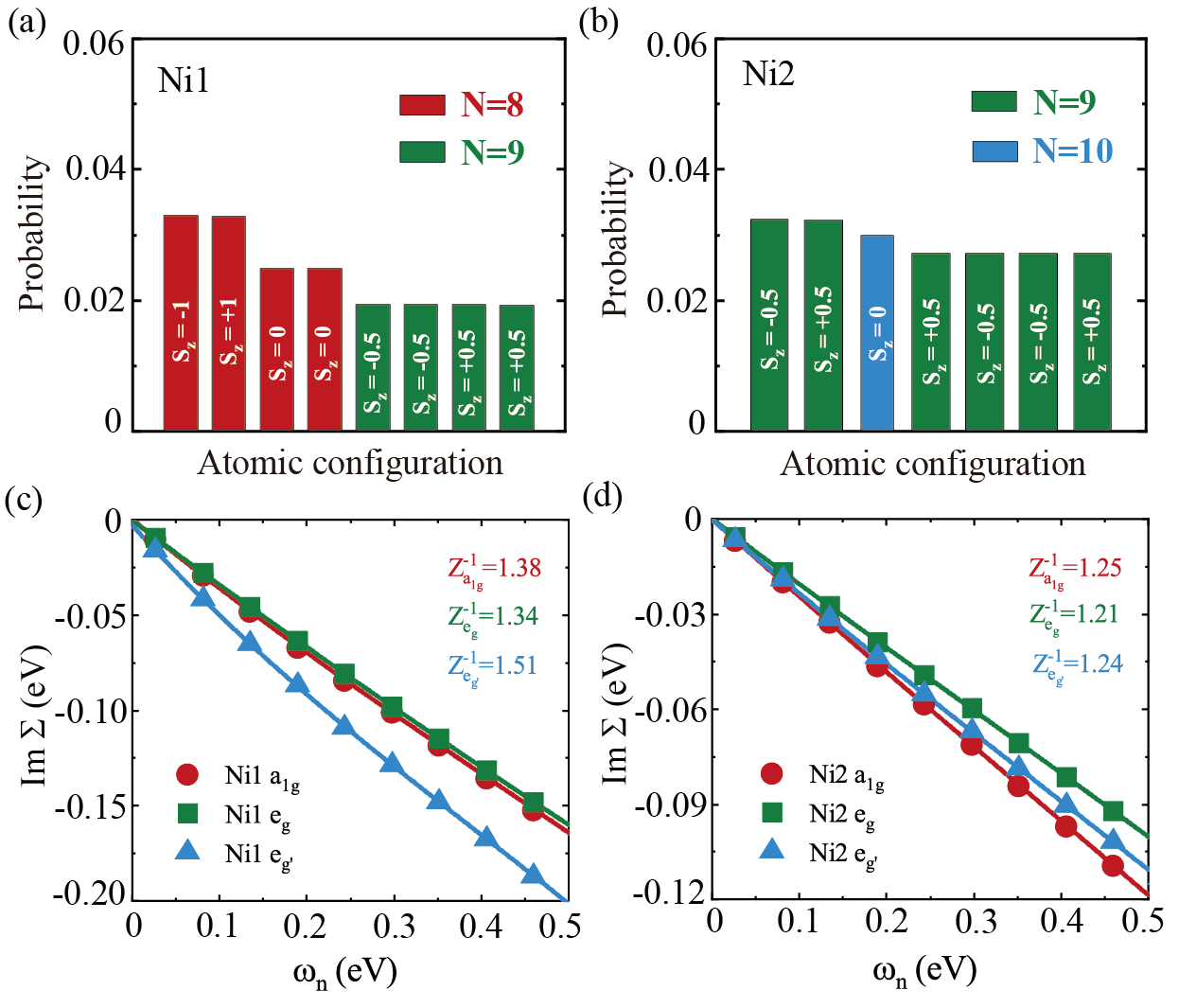}
\caption{(a) The highest probability atomic configurations for Ni1 atoms at 100 K, obtained from the impurity solver in DFT+DMFT calculations. (b) The highest probability atomic configurations for Ni2 atoms at 100 K. (c) The imaginary part of the orbital-dependent self-energy for Ni1 atoms at 100 K, as calculated using DFT+DMFT. Solid lines represent third-order polynomial fits to the data. (d) The imaginary part of the orbital-dependent self-energy for Ni2 atoms at 100 K.}
\label{fig4}
\end{center}
\end{figure}

Due to the correlation effects of Ni-$d$ orbitals, we further investigate the probability of atomic configurations and the self-energy of Ni atoms. Despite Ni$_{3}$GeTe$_{2}$ being a metal with its metallicity primarily derived from Ni-$d$ electrons, well-defined local moments are observable, as illustrated in Figs.~\ref{fig4}(a) and (b). The atomic configuration probabilities reveal that the local moment of Ni1 is larger than that of Ni2, attributable to the lower electron occupation in Ni1. Furthermore, the equal probabilities for spin-up and spin-down configurations lead to a paramagnetic state and significant spin fluctuations, resulting in a broad blurred region in the spectral function of DFT+DMFT calculations compared to DFT results. The comprehensive histograms of atomic configurations are provided in the supplementary information. Figures~\ref{fig4}(c) and (d) present the imaginary part of the self-energy in Matsubara frequency. The band renormalization factor due to correlation effects is derived from $Z^{-1}=1-\partial \text{Im} \Sigma (i\omega)/\partial \omega |_{\omega \rightarrow 0^{+}}$. As shown in Fig.~\ref{fig4}(c) and (d), the renormalization factor ranges from 1.2 to 1.5, indicating weak correlation effects in Ni$_{3}$GeTe$_{2}$. From our results, we propose the physical mechanism in Ni$_{3}$GeTe$_{2}$ involving the factors of layered triangular lattice and the complex magnetic interactions due to bonding with Te and Ge atoms. The flat bands can appear near the Fermi level as a result of the geometry or potential magnetic interaction frustration. And the correlation of $d$-orbitals will further renormalize the flat bands.

We are aware that in narrow-band systems, when the characteristic electronic energy scales become comparable to phonon energies, electron-phonon interactions can be significantly modified and may play a more prominent role \cite{Cao2018,Choi2018}. In recently studied frustrated Kagome lattice systems, electron-phonon interactions have been shown to induce a pronounced kink in the electronic band structure near the Fermi level \cite{Luo2022,Zhong2023}. However, to date, such kink features have rarely been reported in ARPES measurements of  Fe$_{3}$GeTe$_{2}$ and Fe$_{3}$GaTe$_{2}$ \cite{Wu2024,Xu2020}. Moreover, no experimental studies have yet been conducted on Ni$_{3}$GeTe$_{2}$ in this context. In its sister compound Fe$_{3}$GeTe$_{2}$, although electron-phonon interactions have been found to affect transport properties, the electronic band structure shows no significant renormalization in the presence of these interactions \cite{Badrtdinov2023}. Therefore, we recognize that electron-phonon interactions may influence the transport behavior of Ni$_{3}$GeTe$_{2}$, while the associated band renormalization effects are likely to be minimal. Further experimental and theoretical investigations are necessary to fully understand the role of electron-phonon interactions in this system.

Previous studies suggest that, due to different chemical bonding, the antiferromagnetic interaction is strongest in Fe$_{3}$GeTe$_{2}$ \cite{Ruiz2024}. Although the flat band is not closest to the Fermi level in Fe$_{3}$GeTe$_{2}$ \cite{Xu2024}, it may exhibit the largest Kondo coupling due to the significant antiferromagnetic interaction. Consequently, heavy fermion behavior can emerge at lower temperatures, leading to novel quantum states \cite{Zhang2018,Zhao2021}. In Fe$_{3}$GaTe$_{2}$, the flat bands are much closer to the Fermi level \cite{Xu2025}. The stronger ferromagnetic interactions, resulting from the substitution of Ga, stabilize the flat bands and result in an exceptionally high ferromagnetic transition temperature. This enhances the potential for developing advanced spintronic devices. In Ni$_{3}$GeTe$_{2}$, the flat bands are also close to the Fermi level. However, the system is intermediate between ferromagnetic Hund physics and antiferromagnetic Mott physics. As a result, weak magnetic interactions primarily lead to paramagnetism, with only slight renormalization by correlation effects. Additionally, the flat band and VHS in Ni$_{3}$GeTe$_{2}$ are less stable and can be easily influenced by vacancies and external tuning parameters.

At last, we notice that establishing a direct relationship between flat bands and long-range magnetic order remains an open and actively debated question. In certain systems, flat bands are known to enhance Fermi surface instabilities and may promote superconductivity \cite{Luo2022,Zhong2023}. In others, the associated increase in DOS at the Fermi level can enhance the tendency toward magnetic order, potentially satisfying the Stoner criterion for ferromagnetism \cite{Xu2025}. However, superconductivity often competes with long-range magnetic order, making the interplay between these phases highly attractive. Our study indicates that, although a large density of states (DOS) may exist at the Fermi level, the Stoner parameter ($I$) is inherently material-dependent and, to the best of our knowledge, does not have a direct correlation with the presence of flat bands.

\vspace{10pt}
\noindent    
\textbf{Conclusion.} 
In summary, our magnetic DFT+DMFT investigations suggest that the flat bands in Ni$_{3}$GeTe$_{2}$ originate from the geometric frustration of the layered Ni triangular lattice. The significant dependence of magnetic moments on both Coulomb interaction $U$ and Hund's coupling $J$ indicates that the physical behavior of Ni atoms likely falls between Hund and Mott physics. The renormalization effects further influence these flat bands. The sharp peak and dip structure in the density of states (DOS) contributes to the instability of the Fermi surfaces, leading to a rapid decrease in the DOS at the Fermi level due to the effect of Ni vacancy. Through our study, we give a paradigm to understanding the diverse behavior in this type of 2D vdW compounds from the interaction of geometric frustration, the correlation effect and physical behavior of transition metal ions.

\vspace{10pt}
\noindent        
\textbf{Acknowledgements} \\
The authors thank G. E. Volovik for the fruitful suggestion. This work is supported by the National Natural Science Foundation of China (Grant Nos. 12204033, 52371174) and the Young Elite Scientist Sponsorship Program by BAST (Grant No. BYESS2023301). Numerical computations were performed on Hefei advanced computing center.

\end{document}